# FlexNPU: Transparent NPU Virtualization for Dynamic LLM Prefill-Decode Co-location

Jiongjiong Gu[1,*], Jianfeng Wang[1], Zidong Han[2], Yongqiao Wang[1], Pengfei Xia[1], Mingjie Zhang[1], Hong Liu[1], Yuanyi Xia[1], Jiajia Chu[1], Yifeng Tang[1], Hui Zang[1], Xin Yao[1], Qijie Qiu[3], Yuzhao Wang[1], Chuanfei Xu[2], Lin Zhang[1], Zhuonan Lai[2], Hongming Huang[1], Jiawei Qiu[2], Gong Zhang[1], Weipeng Cao[2,3,*], Zhong Ming[2,3,*]

[1]Huawei Technologies Co., Ltd, Shenzhen, China 518129

[2]Guangdong Laboratory of Artificial Intelligence and Digital Economy (Shenzhen), Shenzhen, China 518107

[3]College of Computer Science and Software Engineering, Shenzhen University, Shenzhen, China 518060

## ABSTRACT

Modern AI serving increasingly relies on NPUs for conventional inference and large language model serving. However, current NPU deployments commonly expose physical devices directly to applications, which limits runtime control over scheduling and makes it difficult to adapt execution to phase-level workload behavior. This limitation is particularly evident in LLM serving, where the prefill phase is compute-intensive while the decode phase is often constrained by memory bandwidth and KV-cache accesses. Static prefill-decode (PD) disaggregation reduces phase interference, but can introduce resource imbalance and unnecessary data movement. We present FlexNPU, a transparent user-space virtualization layer for Ascend NPUs. FlexNPU interposes on AscendCL APIs and routes NPU operations through per-device daemons, decoupling unmodified from physical NPU devices without modifying model code, AI frameworks, or NPU drivers. This runtime boundary allows FlexNPU to virtualize NPU objects, control operator dispatch, and support phase-aware scheduling for LLM serving. In particular, FlexNPU enables dynamic PD co-location, which adapts scheduling between prefill and decode according to their complementary resource characteristics. We implement FlexNPU on Huawei Ascend NPUs and evaluate it with typical LLM workloads. Compared with direct NPU passthrough, FlexNPU introduces no measurable inference overhead and slightly improves throughput in some scenarios. On a 384-card Ascend 910C deployment of DeepSeek-R1, FlexNPU improves throughput over static PD disaggregation by 5.15% and 26.33%. On Qwen2.5-7B, compared with static PD co-location, FlexNPU maintains comparable throughput while reducing TTFT by over 92% across tested workloads with nearly unchanged TPOT. These results show that transparent NPU virtualization is a practical substrate for efficient and responsive LLM serving.

## 1. INTRODUCTION

Large language models (LLMs) and modern AI services are reshaping cloud accelerator infrastructure. In addition to conventional computer vision and recommendation inference, cloud platforms increasingly serve LLM workloads with diverse request lengths, dynamic batching behavior, and distinct execution phases. Neural processing units (NPUs), such as Huawei Ascend NPUs, have become an important class of accelerators for these workloads. As NPU clusters scale in both size and cost, improving their serving efficiency has become a critical systems problem.

Today, many NPU deployments expose physical devices directly to applications through the vendor runtime stack. This passthrough model is simple and efficient, but it provides limited opportunity for runtime-level control. Once an application is bound to a physical NPU, the system has little visibility into its execution behavior and limited ability to adjust scheduling decisions without modifying the application, the serving framework, or the NPU driver. This restriction becomes increasingly problematic as AI workloads become more heterogeneous and phase-dependent.

LLM serving is a representative example. A request is processed in two major phases: prefill, which consumes the input prompt and is typically compute-intensive, and decode, which generates tokens autoregressively and is often constrained by memory bandwidth and KV-cache accesses. These two phases stress the accelerator in different ways. Running them together (i.e., static PD co-location) can cause interference, while statically separating them onto different devices can lead to resource imbalance [1]. For example, as shown in Figure 1, prefill workers may leave memory bandwidth underutilized, whereas decode workers may not fully use available compute capacity. Moreover, static PD disaggregation may introduce KV-cache movement and more complex coordination across devices or nodes.

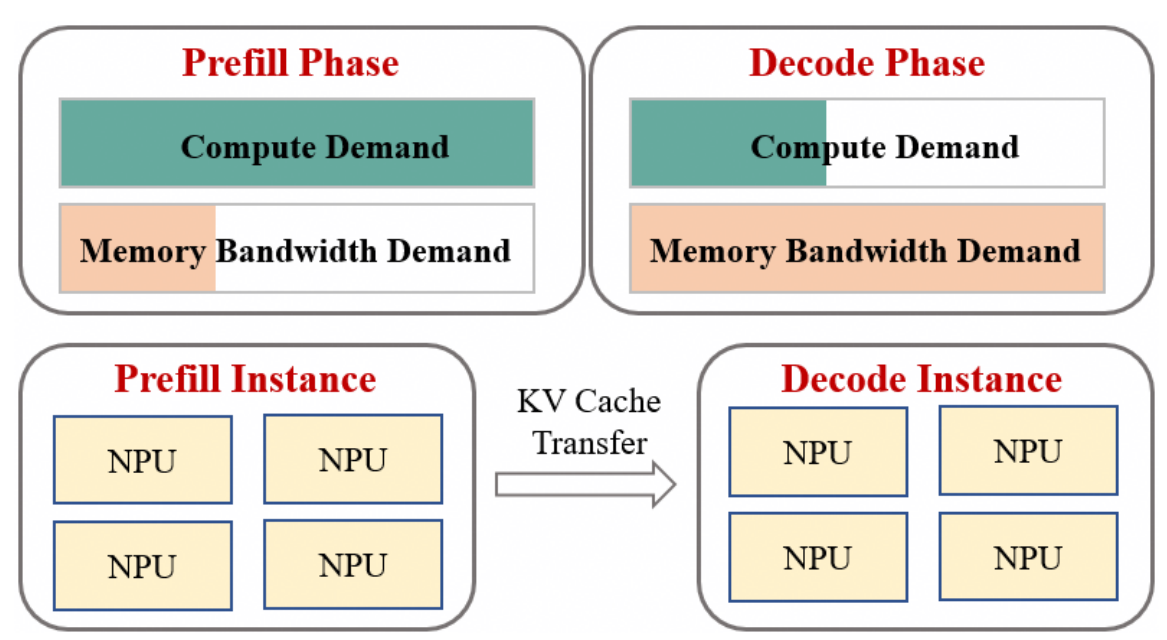

Figure 1. Comparison of compute and bandwidth requirements between prefill and decode phases in LLM inference.

This creates a fundamental trade-off. Co-locating prefill and decode can avoid unnecessary data movement and improve locality, but may suffer from phase interference if scheduling is not carefully controlled. Disaggregating prefill and decode can isolate their resource usage, but may waste resources when the workload mixes changes. Since request distributions, prompt lengths, output lengths, and batching opportunities vary over time, a fixed deployment strategy is often suboptimal. An efficient LLM serving system should be able to adapt how prefill and decode share NPU resources at runtime [2].

A natural place to provide such control is the NPU runtime boundary. On Ascend platforms, AI frameworks and serving systems ultimately interact with devices through AscendCL (Ascend Computing Language), a unified API framework, which provides APIs for device management, memory allocation, stream and event management, synchronization, and model or operator execution. This API layer is narrow enough to intercept, yet low-level enough to cover different frameworks and workloads. Interposing at this boundary allows a system to observe and control NPU operations without requiring changes to application code or NPU drivers.

*Corresponding authors. Emails: dennis.gu@huawei.com (Jiongjiong Gu), caoweipeng123@gmail.com (Weipeng Cao), mingz@szu.edu.cn (Zhong Ming)

This paper presents FlexNPU, a transparent user-space virtualization layer for Ascend NPU serving. FlexNPU intercepts AscendCL APIs using a client library injected into unmodified application processes. Intercepted requests are forwarded to per-device FlexNPU daemons, which manage runtime states, create virtualized execution contexts, and dispatch operations to physical NPUs. This design decouples application-visible NPU resources from physical devices while preserving compatibility with existing model code and serving frameworks.

FlexNPU uses this virtualization layer to support phase-aware LLM serving. Instead of statically choosing between PD co-location and disaggregation, FlexNPU enables dynamic PD co-location. Prefill and decode are executed as separate logical components, but FlexNPU can schedule their NPU operations through a shared runtime layer. This allows FlexNPU to exploit their complementary resource characteristics and adjust the execution balance between the two phases according to workload conditions.

A key requirement for such a design is low overhead. Runtime interposition is only practical if it does not significantly slow down inference. FlexNPU therefore keeps the interception path lightweight: it intercepts a small set of stable AscendCL APIs, virtualizes handles in user space, and forwards requests through efficient communication with per-device daemons. Large tensor data are not copied through the control path. As a result, FlexNPU can preserve the performance benefits of direct NPU execution while enabling additional scheduling control.

We implement FlexNPU on Huawei Ascend NPUs and evaluate it with representative LLM workloads. Our evaluation focuses on three questions. First, does transparent AscendCL interposition introduce overhead compared with direct passthrough? Second, can FlexNPU's dynamic PD co-location improve large-scale LLM serving throughput compared with static PD disaggregation? Third, when prefill and decode are already co-located, can FlexNPU's dynamic scheduling improve serving efficiency and responsiveness compared with static PD co-location?

Our results show that FlexNPU is practical and effective. Compared with direct NPU passthrough, FlexNPU introduces no measurable end-to-end inference overhead in our experiments and slightly improves throughput in several scenarios. For large-scale LLM serving, we deploy DeepSeek-R1 on a 384-card Ascend 910C cluster and compare FlexNPU's dynamic PD co-location with static PD disaggregation. FlexNPU improves system throughput by 5.15% and 26.33% under two representative workload distributions. We further evaluate Qwen2.5-7B by comparing FlexNPU's dynamic PD co-location with static PD co-location. FlexNPU achieves comparable or slightly higher throughput in most tested workloads and dramatically reduces TTFT from hundreds of seconds to sub-second levels, while keeping TPOT nearly unchanged. These results indicate that FlexNPU's PD-aware scheduling benefits are not limited to a single deployment mode or model scale.

This paper makes the following contributions:

(1) **Transparent user-space NPU virtualization.** We design and implement FlexNPU, a user-space virtualization layer that interposes on AscendCL APIs and decouples unmodified AI applications from physical Ascend NPU devices.

(2) **Low-overhead runtime interposition.** We show that AscendCL-level interception can be implemented with negligible end-to-end inference overhead compared with direct passthrough, and can even improve performance in some scenarios.

(3) **Dynamic PD co-location for LLM serving.** We use FlexNPU's virtualization layer to dynamically co-locate and schedule LLM prefill and decode execution, improving throughput over static PD co-location and PD disaggregation on large-scale Ascend NPU deployments.

(4) **Evaluation on real NPU workloads.** We evaluate FlexNPU with representative LLM workloads, including DeepSeek-R1 on 384 Ascend 910C NPUs and Qwen2.5-7B. The evaluation demonstrates that FlexNPU provides transparent virtualization with no measurable inference overhead, improves throughput over static PD disaggregation for large-scale DeepSeek-R1 serving, and substantially reduces TTFT compared with static PD co-location for Qwen2.5-7B.

# 2. RELATED WORK

## 2.1 Transparent Accelerator Virtualization and Runtime Interposition

Cloud AI services increasingly rely on accelerator sharing to improve utilization and reduce deployment cost. A single inference workload often cannot fully utilize the compute, memory, or bandwidth capacity of a modern accelerator. Prior work has therefore explored accelerator virtualization and runtime interposition to enable multi-tenant sharing, isolation, and scheduling.

A large body of work focuses on GPU sharing. gShare [3] decouples GPU resource management from the traditional CPU-centric control path and enables transparent fine-grained sharing through vGPU remapping. It reduces GPU usage by 43%-63% while satisfying most users' latency targets. Nixie [4] provides transparent temporal multiplexing for commodity GPUs by coordinating GPU memory allocation and kernel launches, improving interactive task latency. FaST-GShare [5] proposes a spatio-temporal GPU sharing architecture for FaaS workloads and mitigates memory contention through model sharing. StreamBox [6] designs a lightweight GPU sandbox with fine-grained memory management and in-GPU communication across functions, reducing GPU memory consumption by up to 82%.

Other systems focus on resource pooling and cluster-level scheduling. Singularity [7] describes Microsoft's globally distributed AI scheduling service, which supports transparent preemption, migration, and elastic resizing across GPU and FPGA clusters. Aegaeon [8] enables token-granularity GPU pooling and autoscaling for LLM serving by combining component reuse, explicit memory management, and fine-grained KV-cache synchronization.

These systems demonstrate the value of transparent accelerator management, but they are primarily designed for GPUs and often rely on GPU-specific mechanisms such as CUDA interception, mature GPU memory-management primitives, SM-level scheduling, or MIG-style partitioning. These mechanisms do not directly apply to Ascend NPUs, whose runtime interface and resource-control mechanisms differ substantially from GPUs. FlexNPU targets this gap by interposing on AscendCL, the runtime

API boundary used by Ascend applications. By virtualizing runtime objects and routing operations through user-space per-device daemons, FlexNPU provides transparent NPU virtualization without modifying model code, AI frameworks, or NPU drivers.

## 2.2 Efficient AI Inference on Shared Accelerators

Prior work has also studied how to improve the efficiency of AI inference services on shared accelerator infrastructure. These systems optimize different parts of the stack, including model placement, memory sharing, cold-start reduction, and accelerator provisioning.

Several systems improve resource efficiency through model-aware scheduling and memory coordination. Prism [9] exploits cross-model memory sharing for multi-LLM inference. By combining on-demand memory allocation with a two-level scheduling policy, it achieves more than 2x cost reduction. InSS [10] develops an interference-aware latency model and a two-stage scheduler to jointly optimize model placement and GPU resource allocation, improving throughput by up to 86% over state-of-the-art schedulers. Tetris [11] improves serverless inference density by deduplicating tensors and sharing them at runtime.

Another line of work focuses on reducing startup and model-loading overhead in serverless AI inference. FaaSnap [12] uses VM snapshotting to accelerate FaaS cold starts. CXLfork [13] explores a CXL-based remote fork interface for near-zero-copy process cloning. Optimus [14] reuses structurally similar models in warm containers to reduce model-loading latency. InstaInfer [15] proposes opportunistic preloading to reduce ML model loading overhead. ServerlessLLM [16] targets low-latency serverless LLM inference through fast checkpoint loading and inference migration. Torpor [17] keeps models in host memory and lazily binds them to GPUs when requests arrive, allowing a node to serve many inference functions concurrently.

These systems address important efficiency challenges in AI serving, but their focus differs from FlexNPU. Most of them target GPU-based deployments or serverless cold-start reduction. In contrast, FlexNPU focuses on the runtime boundary of Ascend NPU serving. Its goal is not to redesign model loading or serverless scaling, but to provide a low-overhead transparent virtualization layer that enables additional runtime control over NPU execution. In this paper, we use this substrate to support dynamic PD co-location for LLM serving and evaluate both the overhead of the virtualization layer and its impact on LLM serving throughput.

## 2.3 LLM Serving: PD Disaggregation and PD Co-location

LLM inference consists of two phases with different resource characteristics. The prefill phase processes the input prompt and is typically compute-intensive and highly parallelizable. The decode phase generates tokens autoregressively and is often constrained by memory bandwidth, KV-cache accesses, and limited token-level parallelism. This phase heterogeneity has motivated recent systems to separate, co-locate, or dynamically multiplex prefill and decode execution.

**PD disaggregation.** Splitwise [18] explores separating prefill and decode onto different GPUs to isolate their resource demands. DistServe [19] further optimizes resource allocation for disaggregated prefill and decode execution in heterogeneous clusters. These systems show that disaggregation can reduce phase interference, but static disaggregation can also suffer from resource imbalance: one phase may become the bottleneck while resources assigned to the other phase remain underutilized. semi-PD [1] proposes an intermediate design that asynchronously decouples prefill and decode computation at the SM level while using a unified memory manager to avoid KV-cache transfer and weight replication overheads.

**PD co-location and multiplexing.** Recent systems further explore dynamic co-location within a GPU. MuxServe [20] proposes a spatio-temporal multiplexing system for multi-LLM serving. It co-locates models according to popularity to reuse memory and exploits prefill/decode phase characteristics for flexible co-deployment. MuxWise [2] introduces an intra-GPU prefill-decode multiplexing paradigm with a bubble-free multiplexing engine, an interference-resilient estimator, and an SLO-aware dispatcher. Bullet [21] enables concurrent execution of prefill and decode requests through fine-grained phase coordination and an online performance model. Nexus [22] identifies diminishing returns in GPU resource allocation and memory-bandwidth contention as key bottlenecks, and proposes proactive intra-GPU PD decoupling. DuetServe [23] provides PD isolation within a single GPU by dynamically activating SM-level spatial multiplexing when inter-token latency is predicted to degrade. RouterWise [24] studies joint request routing and resource allocation for multi-model LLM serving in GPU clusters.

These systems demonstrate that exploiting the heterogeneity between prefill and decode is important for efficient LLM serving. However, most existing designs are built on GPU platforms and assume GPU-specific spatial multiplexing or CUDA-level scheduling mechanisms, such as SM partitioning, MIG, or fine-grained CUDA stream control. Such mechanisms are not directly available or equally mature on current Ascend NPU platforms.

FlexNPU takes a different approach. Rather than relying on GPU-specific spatial partitioning, FlexNPU introduces a transparent user-space virtualization layer at the AscendCL boundary. This layer allows FlexNPU to observe and control NPU operations issued by unmodified applications. On top of this substrate, FlexNPU supports dynamic PD co-location for Ascend NPU serving. Compared with static PD co-location and PD disaggregation, FlexNPU can adjust prefill and decode execution through runtime scheduling, reducing resource imbalance while preserving application transparency.

**Remark.** Existing work has extensively studied GPU virtualization, accelerator sharing, serverless AI inference, and LLM prefill/decode scheduling. FlexNPU differs from prior systems in three ways. First, it targets Ascend NPUs rather than GPUs, and therefore operates through AscendCL instead of CUDA or GPU-specific partitioning mechanisms. Second, it provides transparent user-space virtualization, requiring no changes to model code, serving frameworks, or NPU drivers. Third, it uses this virtualization layer to support dynamic PD co-location for LLM serving on NPUs. These design choices make FlexNPU complementary to prior GPU-based LLM serving systems and provide a practical runtime substrate for efficient NPU-based AI serving.

# 3. DESIGN OF FlexNPU

AscendCL is the runtime API layer for developing deep learning applications on Ascend NPUs. It provides APIs for device and context management, memory allocation, model loading and execution, operator invocation, stream and event management, synchronization, and data movement. Modern AI frameworks and serving systems, such as PyTorch, MindSpore, and vLLM-based stacks, eventually interact with Ascend devices through this runtime layer. This makes AscendCL a natural interception boundary for transparent NPU virtualization.

FlexNPU is built on two observations. First, many AI applications access NPUs through a relatively narrow and stable set of runtime APIs. Interposing at this boundary can provide runtime control without modifying model code, AI frameworks, or NPU drivers. Second, LLM inference exhibits strong phase-level heterogeneity. The prefill phase is typically compute-intensive and benefits from parallel execution, whereas the decode phase is often limited by memory bandwidth and KV-cache accesses. As shown in Figure 2, during decode, increasing the number of compute units initially improves memory bandwidth utilization, but the bandwidth quickly approaches saturation. Beyond this point, assigning additional compute resources brings limited throughput improvement. This behavior suggests that static resource assignment can leave either compute or memory bandwidth underutilized.

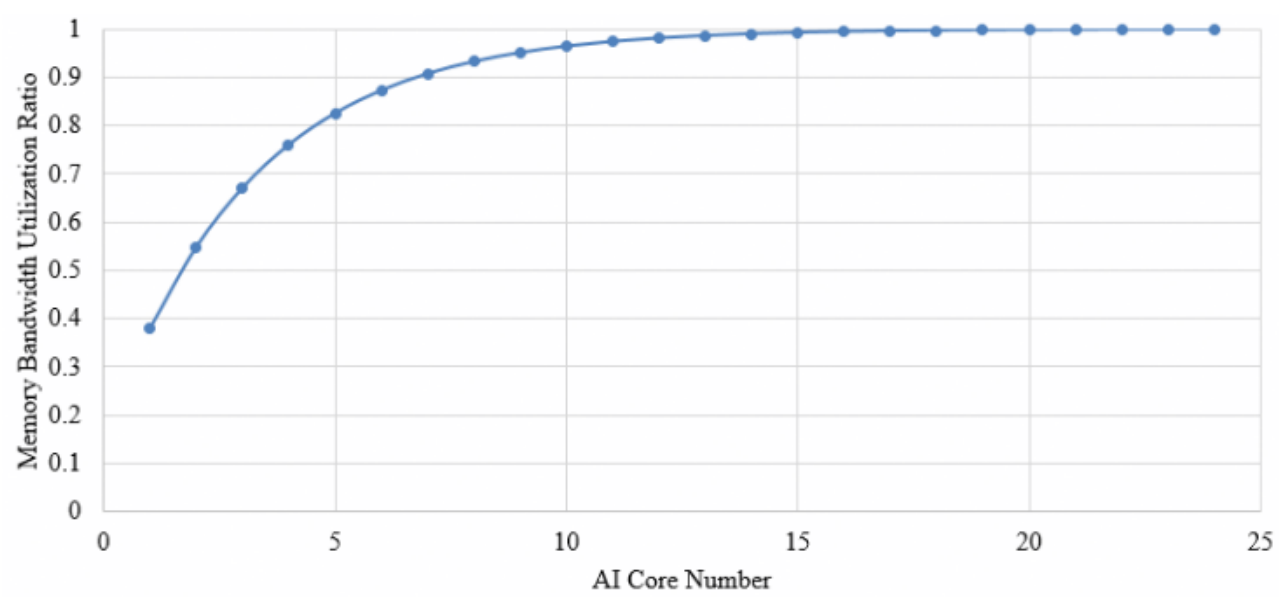

Figure 2. Memory bandwidth utilization under different AI Core allocations during LLM decode.

FlexNPU addresses this mismatch by inserting a transparent user-space virtualization layer between AI applications and physical Ascend NPUs. By intercepting AscendCL calls, FlexNPU decouples application-visible NPU objects from physical resources and routes NPU operations through a runtime proxy. This design enables FlexNPU to observe and control NPU execution at runtime while preserving application compatibility. In this paper, we focus on using this substrate for two purposes: low-overhead transparent NPU virtualization and dynamic prefill/decode co-location for LLM serving.

FlexNPU does not require hardware support for full spatial partitioning of Ascend NPUs. When hardware or runtime mechanisms for stronger isolation are available, FlexNPU can use them as optional enforcement mechanisms. Otherwise, FlexNPU relies on user-space dispatch control, stream scheduling, and phase-aware temporal multiplexing. This design makes FlexNPU deployable on existing Ascend platforms while still enabling runtime scheduling decisions that are difficult to implement in direct passthrough mode.

FlexNPU is designed with three goals:

(1) **Transparency.** Existing applications should run without modifying model code, serving frameworks, or NPU drivers.

(2) **Low overhead.** Runtime interposition should not noticeably slow down inference compared with direct NPU passthrough.

(3) **Phase-aware LLM serving.** FlexNPU should support dynamic scheduling between prefill and decode to reduce the inefficiency of static PD co-location and PD disaggregation.

## 3.1 Overview

Figure 3 shows the architecture of FlexNPU. FlexNPU intercepts AscendCL calls issued by AI applications through common frameworks and serving systems. The intercepted calls are forwarded to a FlexNPU daemon, which virtualizes runtime objects, manages request queues, and dispatches operations to physical NPUs. A lightweight scheduler uses runtime statistics and phase profiles to guide dispatch decisions, especially for LLM PD co-location.

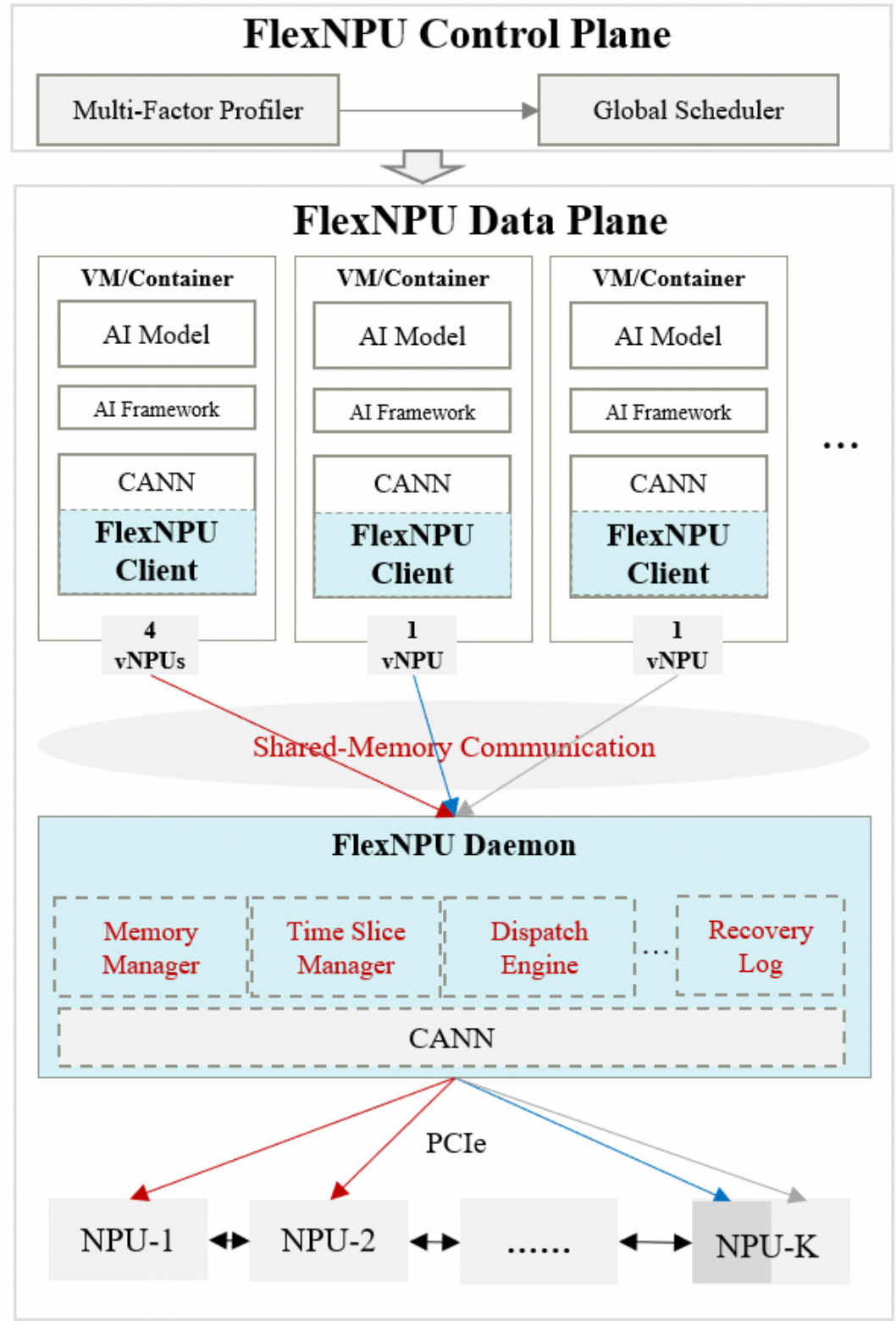

Figure 3. FlexNPU architecture. FlexNPU intercepts AscendCL calls in user space and routes NPU operations through per-device daemons for transparent virtualization and PD-aware scheduling.

FlexNPU consists of three major components:

**(1) FlexNPU Client.** The client library is injected into each application process using *LD_PRELOAD*. It intercepts selected AscendCL APIs, virtualizes application-visible handles, and forwards requests to the FlexNPU daemon. Since the interception happens below the AI framework, applications and model-serving systems do not need to be modified.

**(2) FlexNPU Daemon.** The daemon runs in user space and manages local Ascend NPUs. It receives intercepted requests from one or more client processes, maintains mappings between virtual and physical NPU objects, and dispatches operations to physical devices. The daemon also collects runtime statistics such as operator latency, queueing delay, memory usage, and phase-level execution behavior.

**(3) FlexNPU Scheduler and Profiler.** The scheduler and profiler form the runtime policy module. The profiler records the execution characteristics of different models and phases, including compute demand, memory footprint, and bandwidth pressure. The scheduler uses these profiles and online measurements to decide how to dispatch operations from different instances. For LLM serving, it dynamically adjusts the execution balance between prefill and decode.

FlexNPU follows a two-plane design. The data plane handles latency-sensitive API interception, request forwarding, handle translation, and operator dispatch. It is designed to be lightweight and adds minimal overhead on the critical path. The policy plane performs profiling, runtime monitoring, and scheduling policy updates. This separation allows FlexNPU to make runtime scheduling decisions without slowing down individual AscendCL calls.

A key design principle of FlexNPU is to virtualize resources at the runtime boundary rather than inside the AI framework or the NPU driver. This has three benefits. First, it preserves application transparency. Second, it avoids intrusive changes to vendor drivers or firmware. Third, it exposes enough runtime information, such as memory allocation, stream creation, synchronization, and operator execution, for FlexNPU to perform phase-aware scheduling.

## 3.2 Transparent Interception and Proxying

FlexNPU interposes on AscendCL APIs and redirects NPU operations through a user-space proxy. The intercepted calls include device and context management, memory allocation and deallocation, memory copy, stream and event creation, synchronization, model execution, and operator launch. Examples include APIs such as *aclrtMalloc*, *aclrtFree*, *aclrtCreateStream*, *aclrtCreateEvent*, and model or operator execution interfaces.

FlexNPU chooses low-level and stable AscendCL APIs as interception points. Intercepting at this layer provides broad coverage across different AI frameworks while avoiding framework-specific integration. For example, PyTorch, MindSpore, and LLM serving frameworks may use different execution engines, but their NPU operations eventually go through AscendCL. FlexNPU therefore only needs to intercept a small set of runtime APIs to support a wide range of applications. This boundary also keeps FlexNPU independent of model structure. FlexNPU does not need to understand the full computation graph or modify model operators. Instead, it observes runtime-level operations and controls how these operations are forwarded to physical NPUs.

When an application invokes an intercepted AscendCL API, the FlexNPU client packages the request into a compact descriptor and forwards it to the daemon. The descriptor contains the operation type, virtual handles, metadata, and necessary arguments. Large payloads, such as tensor data, are not copied through the control path. Instead, FlexNPU forwards metadata and manages device memory through virtualized handles. The daemon validates each request, translates virtual objects to physical resources, and inserts the operation into the corresponding dispatch queue. Depending on the current policy, the daemon may dispatch the operation immediately or delay it briefly to coordinate execution with other operations. This design allows FlexNPU to control execution order without introducing expensive data movement. To reduce critical-path overhead, FlexNPU keeps the forwarding path lightweight. The client and daemon communicate through efficient shared-memory channels and avoid heavyweight RPC mechanisms. Most intercepted calls require only handle translation and metadata forwarding. As a result, FlexNPU can preserve the performance characteristics of direct NPU execution.

AscendCL exposes runtime objects such as devices, contexts, streams, events, and memory pointers. Directly exposing physical handles to applications would tightly bind an application process to a specific device and make runtime scheduling difficult. FlexNPU therefore introduces virtual handles. Each application sees a set of virtual NPU objects. The daemon maintains the mapping from virtual handles to physical resources. For example, a virtual stream is mapped to a physical stream managed by the daemon, and a virtual memory object is mapped to a physical NPU memory allocation. Applications continue to use standard AscendCL semantics, while FlexNPU controls the underlying resource mapping. Handle virtualization is essential for transparency. It allows FlexNPU to interpose on NPU operations without changing application behavior. It also gives the daemon a consistent view of streams, events, memory objects, and execution requests, which is necessary for runtime scheduling.

A major concern for runtime interposition is overhead. FlexNPU applies several optimizations to keep overhead low. First, FlexNPU intercepts only the AscendCL APIs needed for execution control and resource tracking. It does not interpose on high-level framework logic. Second, FlexNPU avoids copying tensor data through the daemon. The data path remains on device memory or framework-managed host memory, while the daemon handles only metadata and runtime control. Third, FlexNPU batches lightweight bookkeeping operations when possible and reuses virtual-to-physical mappings to avoid repeated lookup overhead. These optimizations make FlexNPU suitable for latency-sensitive inference workloads. As shown in Section 4, FlexNPU introduces no measurable end-to-end inference overhead in our experiments compared with direct passthrough, and can improve performance in some scenarios.

## 3.3 Runtime Profiling

FlexNPU uses lightweight runtime profiling to understand the execution characteristics of models and LLM phases. The profiler collects statistics from the interception layer and the daemon, including operator execution time, queueing delay, memory allocation size, stream usage, and phase-level throughput.

For LLM serving, FlexNPU profiles prefill and decode separately. Prefill typically has high compute demand and benefits from additional parallel execution resources. Decode, in contrast, tends to become memory-bandwidth-bound as the number of active sequences grows. Figure 2 illustrates this behavior: during decode, memory bandwidth utilization increases with the number of compute units at first, but eventually saturates. After saturation, assigning more compute units provides limited benefit.

These profiles are not intended to be perfect offline models. Instead, they provide coarse but useful guidance for runtime scheduling. FlexNPU continuously updates statistics during execution and uses them to adjust the scheduling balance between prefill and decode. This lightweight approach avoids expensive profiling procedures while still capturing the phase-level behavior needed for dynamic co-location.

## 3.4 Dynamic PD Co-Location for LLM Serving

As shown in Figure 1, LLM inference exposes a fundamental resource imbalance between prefill and decode. The prefill phase processes input prompts and is typically compute-intensive. The decode phase generates tokens sequentially and is often limited by memory bandwidth and KV-cache accesses. Assigning dedicated NPU resources to each phase can therefore lead to underutilization: prefill workers may underuse memory bandwidth, while decode workers may underuse compute capacity after bandwidth saturation.

Existing deployments often choose between two static strategies. Colocated execution runs prefill and decode on the same worker, which avoids KV-cache transfer but may introduce phase interference. PD disaggregation separates prefill and decode onto different workers or devices, which reduces interference but can introduce KV-cache movement, weight replication, and resource imbalance. Since workload characteristics vary over time, a fixed strategy is often suboptimal.

As shown in Figure 4, FlexNPU enables dynamic PD co-location on Ascend NPUs. Prefill and decode are executed as separate logical instances, but their NPU operations (e.g., AI cores management) are routed through the FlexNPU daemon. The daemon maintains separate queues for prefill and decode requests and dynamically adjusts their dispatch ratio according to runtime conditions.

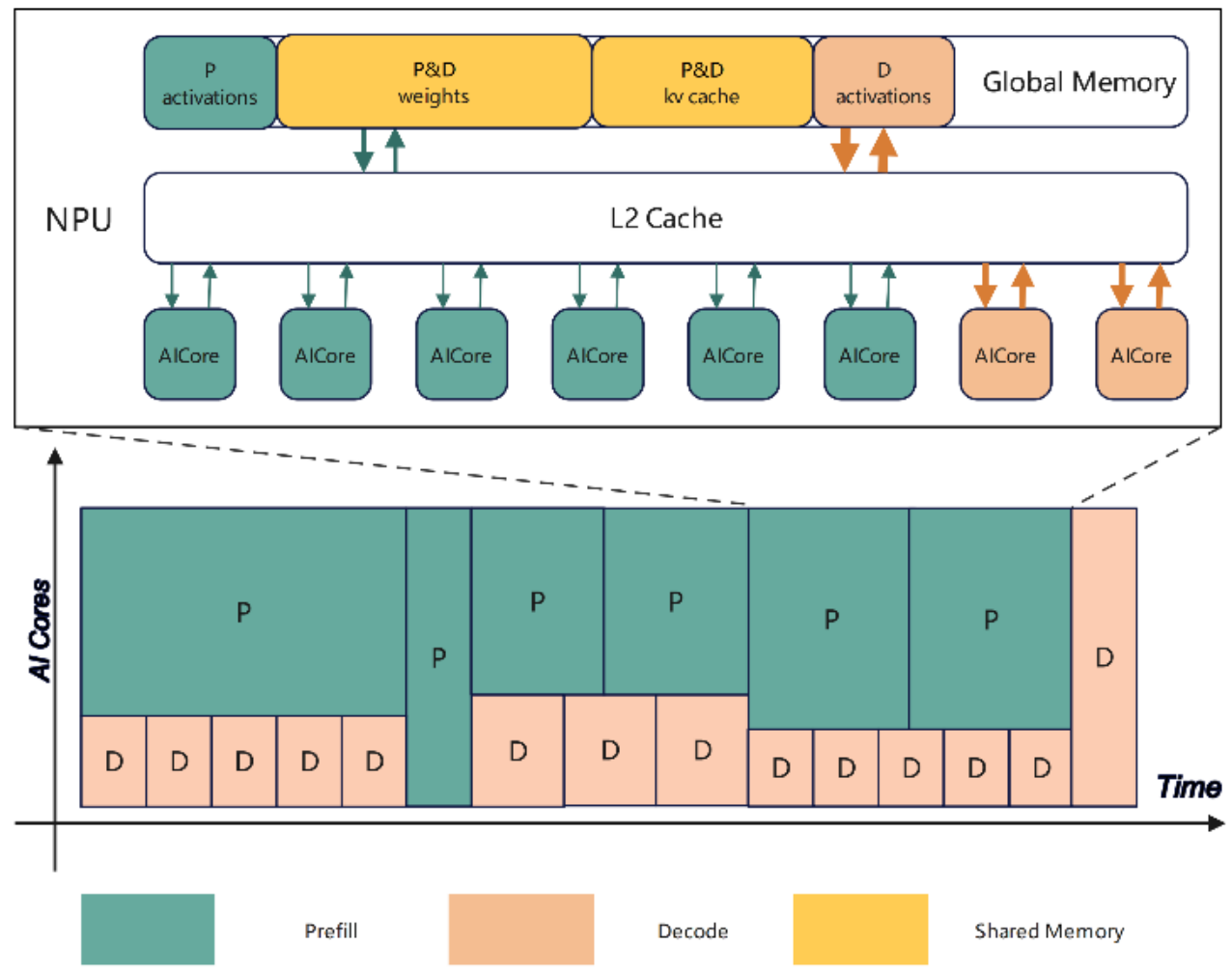

Figure 4. Dynamic PD co-location schematic.

**Phase-Aware Dispatch:** FlexNPU classifies requests by phase and schedules them through phase-aware queues. At runtime, the daemon monitors: (1) the number of pending prefill and decode operations; (2) recent prefill and decode execution times; (3) observed memory bandwidth pressure; (4) decode progress and active sequence count; and (5) NPU queue occupancy and device utilization.

Based on these signals, FlexNPU adjusts the dispatch ratio between prefill and decode. When decode dominates and memory bandwidth is close to saturation, giving decode more compute slots may not improve throughput. FlexNPU can then dispatch more prefill work to use otherwise idle compute resources. Conversely, when there is a large decode backlog, FlexNPU increases the decode dispatch share to prevent decode from becoming the serving bottleneck.

This policy does not require hardware-level spatial partitioning. FlexNPU controls the effective execution share of each phase through user-space dispatch decisions and stream scheduling. If platform mechanisms such as stream priorities or resource groups are available, FlexNPU can use them as optional enforcement mechanisms, but the design does not depend on them.

**Avoiding Static PD Imbalance:** Static PD co-location and PD disaggregation requires operators to choose a fixed resource split between prefill and decode. This split is difficult to tune because the best ratio depends on prompt length, output length, batching behavior, and request arrival patterns. A ratio that works well for one workload may underutilize resources under another.

FlexNPU avoids this rigidity by adjusting the prefill/decode balance online. For workloads with short inputs and long outputs, decode occupies a larger fraction of execution time, and FlexNPU can allocate more dispatch opportunities to decode while still using available compute gaps for prefill. For workloads where input and output lengths are more balanced, FlexNPU can exploit the complementary resource demands of the two phases more effectively, leading to larger throughput gains.

**Memory Locality:** Dynamic co-location also improves locality compared with fully disaggregated execution. When prefill and decode are placed within the same NPU pool, FlexNPU can reduce unnecessary KV-cache movement between disaggregated workers. This avoids part of the communication and coordination overhead associated with static PD separation.

FlexNPU does not require applications to implement new data movement logic. Since prefill and decode operations are routed through the same runtime layer, FlexNPU can coordinate their execution while preserving the application-visible AscendCL interface.

## 3.5 Discussion and Scope

FlexNPU is designed as a runtime virtualization substrate for Ascend NPU serving. In this paper, we focus on two capabilities that are implemented and evaluated: transparent AscendCL-level virtualization and dynamic PD co-location for LLM serving.

FlexNPU does not require modifications to application code, AI frameworks, or NPU drivers. It also does not assume full hardware support for spatial partitioning. This makes FlexNPU practical for existing Ascend deployments. At the same time, FlexNPU can benefit from stronger hardware mechanisms when they are available, such as stream priorities, resource groups, or future NPU partitioning support.

Other capabilities, such as full SLO-aware autoscaling, transparent live migration, and failure recovery, are complementary to FlexNPU's virtualization substrate but are outside the scope of this paper. We leave their full design and evaluation to future work.

# 4. EVALUATION

We evaluate FlexNPU to answer the following questions:

**(1) Virtualization overhead.** Does FlexNPU introduce noticeable overhead compared with direct NPU passthrough?

**(2) Dynamic PD co-location vs. static PD disaggregation.** Can FlexNPU improve large-scale LLM serving throughput by dynamically co-locating prefill and decode compared with static PD disaggregation?

**(3) Dynamic PD co-location vs. static PD co-location.** When prefill and decode are already co-located on the same NPU resources, can FlexNPU's phase-aware scheduling improve serving responsiveness and maintain throughput compared with static PD co-location?

Our evaluation focuses on two implemented and measured capabilities: transparent AscendCL-level virtualization and dynamic PD co-location. We first quantify the performance impact of FlexNPU compared with native passthrough execution. We then evaluate FlexNPU's PD-aware scheduling on a large-scale DeepSeek-R1 deployment with 384 Ascend 910C NPUs, where the baseline is static PD disaggregation. Finally, Section 4.4 evaluates

Qwen2.5-7B against static PD co-location to study whether FlexNPU can improve co-located serving behavior for a smaller dense LLM.

## 4.1 Experimental Setup

We evaluate FlexNPU on Huawei Ascend 910C NPUs. Our large-scale LLM serving experiments use a CloudMatrix384 supernode with 384 Ascend 910C NPU cards. Each NPU runs the standard Ascend software stack and CANN runtime. Unless otherwise stated, all baselines and FlexNPU use the same NPU driver, CANN version, model implementation, and serving framework.

FlexNPU is implemented as a user-space virtualization layer above AscendCL. The FlexNPU client library is injected into application processes through *LD_PRELOAD* and intercepts selected AscendCL APIs for device management, memory management, stream/event management, synchronization, and model/operator execution. Intercepted requests are forwarded to the FlexNPU proxy daemon, which translates virtualized handles and dispatches operations to physical NPUs according to the scheduling policy.

We use three sets of workloads.

First, to measure the performance impact of FlexNPU's transparent virtualization, we run single-model inference with DeepSeek-R1-Distill-Llama-8B. The model is served by vLLM on Ascend NPUs and evaluated using AISBench with the gsm8k_gen_0_shot_cot_str_perf dataset.

Second, to evaluate dynamic PD co-location at scale, we deploy DeepSeek-R1, a large MoE model, on 384 Ascend 910C NPUs using W8A8 quantization. We evaluate two representative request distributions: (1) 1K-1K: input and output lengths are both around 1K tokens. This workload creates relatively balanced pressure between prefill and decode, and prefill can become the throughput bottleneck. (2) 1K-4K: input length is around 1K tokens and output length is around 4K tokens. This workload is decode-heavy and stresses the decode phase.

Third, we evaluate Qwen2.5-7B in Section 4.4. Unlike DeepSeek-R1, Qwen2.5-7B is a smaller dense LLM. This experiment compares FlexNPU dynamic PD co-location with static PD co-location to study whether FlexNPU improves serving responsiveness and preserves throughput when prefill and decode already share the same NPU resources.

We compare FlexNPU with different baselines depending on the experiment.

(1) Native passthrough. Applications directly use physical NPUs through AscendCL without FlexNPU interception. This baseline is used to quantify the overhead of FlexNPU's transparent virtualization layer.

(2) Static PD disaggregation. Prefill and decode are deployed on separate NPU resources with a fixed resource split. This baseline is used in the DeepSeek-R1 experiment and represents a common deployment strategy for reducing prefill/decode interference in large-scale LLM serving.

(3) Static PD co-location. Prefill and decode are placed on the same NPU resources with a fixed execution policy, without FlexNPU's process-level separation or runtime phase-aware resource adjustment. This baseline is used in the Qwen2.5-7B experiment and represents a common co-located deployment where prefill and decode share accelerator resources but cannot dynamically rebalance execution according to phase-level contention.

We compare these baselines with FlexNPU dynamic PD co-location, where prefill and decode operations are routed through FlexNPU and scheduled dynamically according to phase-level runtime behavior.

We report the following metrics:

(1) End-to-end serving throughput, measured in generated tokens per second or requests per second depending on the experiment.

(2) Relative throughput improvement, normalized to the corresponding baseline.

(3) TTFT, time to first token, which captures prefill responsiveness and user-perceived startup latency.

(4) TPOT, time per output token, which captures decode efficiency and token generation latency.

For the DeepSeek-R1 large-scale serving experiment, we evaluate throughput under deployment latency targets of TTFT$\leqslant$1s and TPOT$\leqslant$50ms. For the Qwen2.5-7B co-location experiment, we report TTFT and TPOT to characterize responsiveness under backlog scenarios and to verify whether throughput changes come at the cost of degraded token-generation latency.

## 4.2 Performance Impact of FlexNPU Virtualization

We first evaluate whether FlexNPU's transparent interception and proxying mechanism introduces overhead compared with direct NPU passthrough. This experiment compares two configurations:

(1) Native passthrough: the application directly invokes AscendCL APIs and uses physical NPUs.

(2) FlexNPU proxy: the application runs unmodified, while AscendCL calls are intercepted and forwarded through FlexNPU.

We use DeepSeek-R1-Distill-Llama-8B as the test model and serve it with vLLM. The vLLM server uses tensor parallelism degree 1, maximum sequence number 512, maximum model length 2048, maximum batched tokens 10240, and NPU memory utilization target 0.9. We use AISBench to generate requests with the gsm8k_gen_0_shot_cot_str_perf dataset. The request rate is 10, the maximum output length is 1024, and the batch size is 384. Generation uses temperature 0.5, top-k 10, top-p 0.95, seed 1214, and repetition penalty 1.03. Each configuration is repeated five times, and we report the average total token throughput.

Table 1 shows the total token throughput under native passthrough and FlexNPU proxy execution.

Table 1. Performance impact of FlexNPU virtualization.

| Configuration | Total token throughput | Relative performance |
|---|---|---|
| Native passthrough | 977.6928 tokens/s | 1.0000x |
| FlexNPU proxy | 988.2675 tokens/s | 1.0108x |

From Table 1, we can observe that: FlexNPU does not reduce inference throughput. Instead, in this experiment, the total token throughput slightly increases from 977.6928 tokens/s to 988.2675 tokens/s, corresponding to a 1.08% improvement over native passthrough.

This result indicates that FlexNPU's user-space virtualization does not introduce noticeable overhead on the end-to-end inference path. The main reason is that FlexNPU intercepts only a small set of low-level AscendCL APIs and does not copy tensor data through the proxy. Large tensors remain in framework-managed host memory or NPU device memory, while FlexNPU forwards only metadata and virtualized handles.

The slight performance improvement comes from FlexNPU's asynchronous proxying design. When the client issues asynchronous AscendCL operations, FlexNPU can return control to the inference worker quickly, while the proxy server transparently handles the asynchronous interaction with the NPU backend. This reduces part of the waiting time in the inference worker and slightly improves computation/communication overlap.

## 4.3 Dynamic PD Co-Location vs. Static PD Disaggregation

We next evaluate whether FlexNPU improves LLM serving efficiency by dynamically co-locating prefill and decode. This experiment uses DeepSeek-R1 deployed on 384 Ascend 910C NPUs.

DeepSeek-R1 is deployed with W8A8 quantization on a CloudMatrix384 supernode. The static PD disaggregation baseline uses a 6P2D deployment: six prefill instances and two decode instances. Each prefill instance uses 16 NPU cards with DP32-EP32 parallelism, while each decode instance uses 144 NPU cards with DP288-EP288 parallelism. This configuration uses all 384 NPU cards.

We compare static PD disaggregation with FlexNPU dynamic PD co-location. Static PD disaggregation assigns fixed resources to prefill and decode. FlexNPU, in contrast, allows prefill and decode to share NPU execution opportunities through runtime scheduling. FlexNPU adjusts the time-slice ratio between prefill and decode according to the workload and phase-level throughput behavior.

All experiments are conducted under the same latency constraints: TTFT $\leqslant$ 1s and TPOT $\leqslant$ 50ms. We evaluate two request distributions: (1) 1K-1K: balanced input/output lengths. In this workload, prefill is the throughput bottleneck under the static PD configuration. (2) 1K-4K: short-input/long-output workload. In this workload, decode is the throughput bottleneck.

Table 2 summarizes the observed prefill and decode bottlenecks under the two request distributions. From Table 2, we can observe that: For the case 1K-1K, prefill becomes the throughput bottleneck. Prefill memory usage is about 56%. Under matched prefill throughput, decode requires at least 25% time-slice share and up to 25% memory usage. While for the case 1K-4K, decode becomes the throughput bottleneck. Decode memory usage is about 39%. Under matched decode throughput, prefill requires up to 40% time-slice share and up to 56% memory usage.

Table 2. Phase-level bottlenecks under different input/output distributions.

| Input/ Output Length | Throughput (RPS) | | | Peak Time Slice Redundancy | | Peak Memory Redundancy | |
|---|---|---|---|---|---|---|---|
| | total | 6P Peak | 2D Peak | Prefill | Decode | Prefill | Decode |
| 1K/1K | 490 | 490 | 812 | N/A | 75% | 44% | 77% |
| 1K/4K | 53 | 490 | 53 | 40% | N/A | 44% | 61% |

These results show that the best PD resource ratio depends on the request distribution. A fixed PD split can be inefficient when the workload shifts. For 1K-1K, prefill becomes the limiting phase, and decode resources can be underutilized. For 1K-4K, decode dominates the serving time, but prefill still requires enough execution opportunities to sustain the decode pipeline.

Figures 5 and 6 further illustrate the relationship between time-slice allocation, throughput, and memory usage.

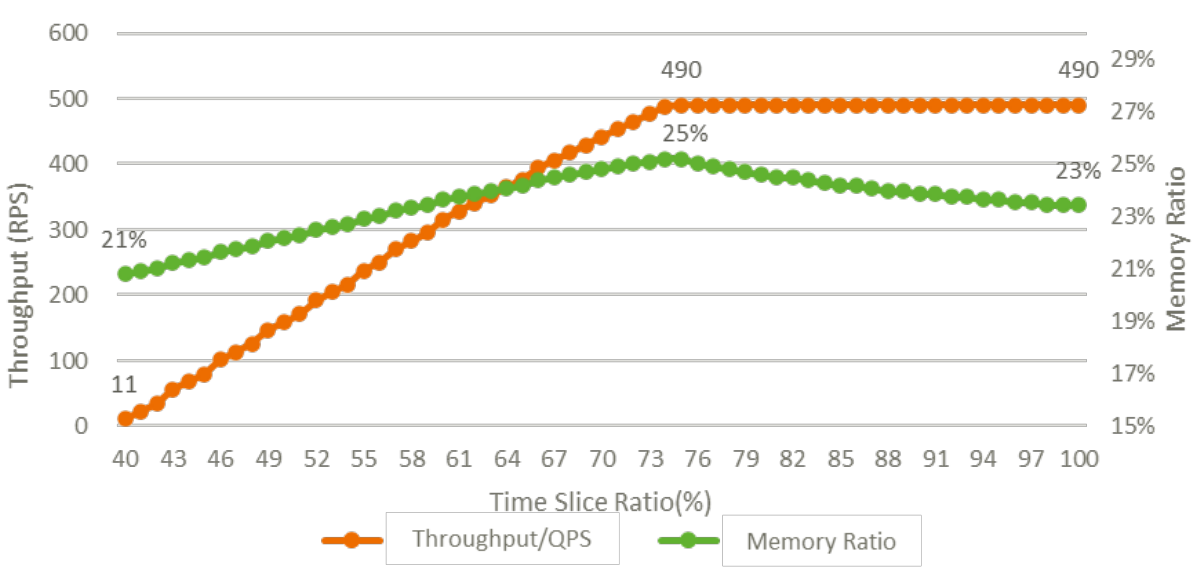

Figure 5. Throughput and memory usage under different decode time-slice ratios for the 1K-1K workload.

From Figure 5, we can observe that, when matching prefill throughput, increasing the decode time-slice ratio first increases system throughput nearly linearly, but the throughput eventually saturates. Memory usage also increases initially and then decreases after reaching a peak, likely because some requests complete generation and release KV-cache memory.

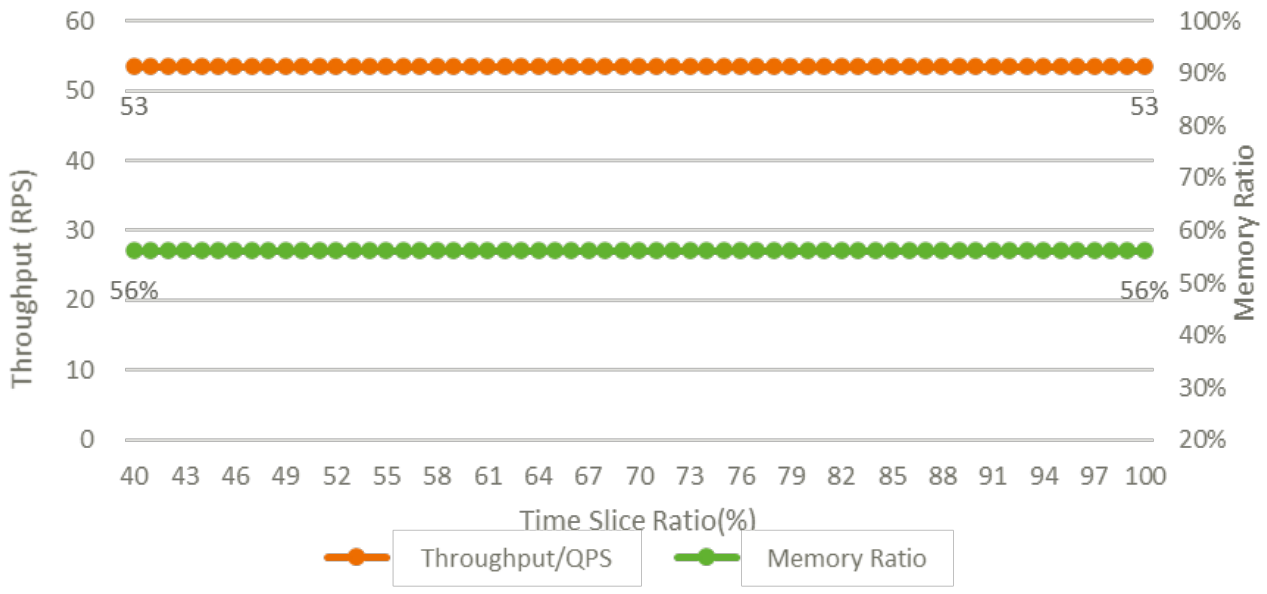

Figure 6. Throughput and memory usage under different prefill time-slice ratios for the 1K-4K workload.

From Figure 6, we can observe that, when matching decode throughput, increasing the prefill time-slice ratio has little impact on throughput and memory usage once decode becomes the dominant bottleneck.

These observations motivate FlexNPU's dynamic PD co-location: instead of using a fixed resource split, FlexNPU adjusts the PD execution ratio according to the current workload bottleneck.

Table 3. DeepSeek-R1 throughput under static PD disaggregation and FlexNPU dynamic PD co-location.

| Deployment mode | Number of instances | Number of NPUs/instance | Input Length | Output Len | Prefill throughput (requests/s) | Decode throughput (requests/s) | Total throughput | Improvement ratio |
|---|---|---|---|---|---|---|---|---|
| PD Disaggregation | 6P2D | P16-D144 | 1024 | 1024 | 489.84 | 811.52 | 489.84 | |
| | | | 1024 | 4096 | 489.84 | 146.63 | 146.63 | |
| PD co-location | 3 | 128 | 1024 | 1024 | 619.18 | 618.18 | 618.18 | 26.33% |
| | | | 1024 | 4096 | 154.4 | 154.18 | 154.18 | 5.15% |

Table 4. Qwen2.5-7B performance under static PD co-location and FlexNPU dynamic PD co-location.

| Deployment mode | Input Length | Output Length | #Requests | Output token throughput (tokens/s) | TTFT (ms) | TPOT (ms) | Throughput Improvement | TTFT Reduction | TPOT Reduction |
|---|---|---|---|---|---|---|---|---|---|
| Static PD co-location | 256 | 256 | 200 | 186.305 | 109941.5 | 20.9980 | — | — | — |
| | 256 | 1024 | 200 | 195.275 | 488099.0 | 20.3510 | — | — | — |
| | 1024 | 256 | 200 | 177.310 | 118164.5 | 21.8010 | — | — | — |
| | 1024 | 1024 | 200 | 189.215 | 506536.5 | 20.8925 | — | — | — |
| FlexNPU dynamic PD co-location | 256 | 256 | 200 | 192.950 | 331.0 | 20.5530 | 3.57% | -99.70% | -2.12% |
| | 256 | 1024 | 200 | 194.250 | 331.5 | 20.5340 | -0.52% | -99.93% | 0.90% |
| | 1024 | 256 | 200 | 183.270 | 8568.5 | 21.5550 | 3.36% | -92.75% | -1.13% |
| | 1024 | 1024 | 200 | 189.515 | 8311.5 | 21.0415 | 0.16% | -98.36% | 0.71% |

Table 3 compares the end-to-end serving throughput of static PD disaggregation and FlexNPU dynamic PD co-location. Under the 1K-4K short-input/long-output workload, FlexNPU improves system throughput by 5.15% over static PD disaggregation. This workload is decode-heavy, and decode already occupies a large fraction of serving time. Therefore, the opportunity for additional prefill/decode balancing is relatively limited. Nevertheless, FlexNPU still improves throughput by reducing the rigidity of the fixed PD split and using available execution gaps more effectively.

Under the 1K-1K balanced workload, FlexNPU improves throughput by 26.33%. In this setting, both prefill and decode contribute significantly to serving execution, and the static PD split suffers from stronger resource imbalance. When one phase becomes temporarily bottlenecked, resources assigned to the other phase may remain underutilized. FlexNPU mitigates this inefficiency by dynamically adjusting the time-slice ratio between prefill and decode.

These results show that static PD disaggregation is not always optimal. The best resource split between prefill and decode depends on input length, output length, phase throughput, and runtime memory behavior. FlexNPU adapts this split at runtime and therefore achieves higher overall throughput.

## 4.4 Dynamic PD Co-Location vs. Static PD Co-Location

We further evaluate FlexNPU using Qwen2.5-7B to understand whether dynamic PD co-location is beneficial beyond the large-scale DeepSeek-R1 deployment. Unlike DeepSeek-R1, which is a large MoE model deployed on hundreds of NPUs, Qwen2.5-7B is a smaller dense LLM. This experiment studies a different deployment scenario: instead of comparing against static PD disaggregation, we compare FlexNPU with static PD co-location, where prefill and decode already share the same NPU resources but do not have FlexNPU's runtime phase-aware scheduling.

We deploy Qwen2.5-7B under two PD deployment strategies: (1) Static PD co-location: prefill and decode are co-located on the same NPU resources and share compute capacity under a fixed execution policy. There is no process-level separation or dynamic resource adjustment between the two phases. (2) FlexNPU dynamic PD co-location: prefill and decode run as separate processes on the same NPU resources. FlexNPU routes their NPU operations through the virtualization layer and dynamically adjusts their effective AI Core execution share according to runtime phase behavior.

We evaluate four input/output length pairs: 256/256, 256/1024, 1024/256, and 1024/1024, covering balanced, decode-heavy, prefill-heavy, and long-context workloads. Each configuration is tested twice with request_rate=4 and max_num_seqs=4, and the average results are reported in Table 4. Given that the request arrival rate equals the maximum number of concurrent sequences, the system experiences request queuing and overload, allowing us to evaluate backlog scenarios where queuing delays stress the scheduler. We measure serving throughput under the same model implementation and serving framework. We also monitor TTFT and TPOT to ensure that throughput improvements do not come from violating latency constraints.

Table 4 compares Qwen2.5-7B under static PD co-location and FlexNPU dynamic PD co-location. FlexNPU achieves comparable or slightly higher throughput than static PD co-location in three out of four workloads, with improvements of 3.57%, 3.36%, and 0.16%. In the decode-heavy 256/1024 workload, FlexNPU's throughput is marginally lower by 0.52%, which is within experimental noise.

The more significant benefit is latency responsiveness. Static PD co-location can suffer from severe head-of-line blocking when prefill and decode contend for AI Core resources under backlog. In our experiments, this leads to extremely high TTFT, reaching hundreds of seconds in some cases. FlexNPU dramatically reduces TTFT across all workloads, bringing it from hundreds-of-seconds levels to sub-second levels. For example, in the 256/1024 workload, TTFT decreases from 488,099 ms to 331.5 ms, a reduction of over 99%. Meanwhile, TPOT remains nearly unchanged, with variations within approximately ±3%.

These results show that static co-location alone is insufficient. Although it avoids the explicit resource split and data movement of PD disaggregation, it does not provide phase-aware control. When prefill and decode compete for the same AI Core resources, static co-location can allow one phase to block the other, causing severe TTFT inflation. FlexNPU avoids this problem by separating prefill and decode at the process level and dynamically adjusting their execution balance through the runtime virtualization layer.

Compared with the DeepSeek-R1 experiment, the absolute throughput improvement on Qwen2.5-7B is smaller because the model size, memory, batching behavior, and prefill/decode compute ratio are different. Nevertheless, the qualitative result is consistent: FlexNPU's dynamic PD co-location improves the serving behavior of static deployments by adapting execution to phase-level demand. For Qwen2.5-7B, the primary benefit is not raw throughput, but a substantial reduction in TTFT while preserving throughput and TPOT.

# 5. CONCLUSION

This paper presented FlexNPU, a transparent user-space virtualization layer for Ascend NPU-based AI serving. FlexNPU is motivated by the limited runtime control provided by direct NPU passthrough and the phase-level resource imbalance in LLM serving. By interposing on AscendCL APIs, FlexNPU decouples unmodified applications from physical NPU devices without requiring changes to model code, serving frameworks, or NPU drivers.

FlexNPU provides a lightweight runtime substrate for NPU execution control. It virtualizes AscendCL runtime objects, routes NPU operations through user-space daemons, and collects runtime information for scheduling. On top of this substrate, FlexNPU supports dynamic PD co-location for LLM serving. Instead of relying on a fixed PD split, FlexNPU adjusts the execution balance between compute-intensive prefill and memory-bandwidth-sensitive decode according to runtime workload behavior.

Our evaluation shows that FlexNPU is practical and effective. Compared with direct NPU passthrough, FlexNPU introduces no measurable inference overhead in our experiments and slightly improves throughput in the DeepSeek-R1-Distill-Llama-8B case due to its asynchronous proxying design. On a 384-card Ascend 910C deployment of DeepSeek-R1, FlexNPU improves serving throughput by 5.15% and 26.33% over static PD disaggregation under two representative workload distributions. We further evaluate Qwen2.5-7B against static PD co-location. FlexNPU achieves comparable or slightly higher throughput in most tested workloads and reduces TTFT by over 92% across all workloads, while keeping TPOT nearly unchanged.

These results demonstrate that transparent runtime-level virtualization is a practical approach for improving NPU-based LLM serving. FlexNPU not only provides low-overhead AscendCL-level interposition, but also enables phase-aware scheduling that addresses different limitations of static PD deployments: it improves throughput over static PD disaggregation in large-scale DeepSeek-R1 serving and substantially improves responsiveness over static PD co-location in Qwen2.5-7B serving. Future work includes extending FlexNPU to support broader multi-tenant scheduling policies, deeper integration with hardware resource isolation mechanisms, and evaluation across more AI serving workloads.